\documentclass[11pt]{iopart}
\usepackage{float}
\usepackage[switch]{lineno}
\usepackage[bottom=0.85in]{geometry}
\usepackage{graphicx}
\usepackage{dcolumn}
\usepackage{bm}
\usepackage{url}
\usepackage[sort&compress,square, comma, numbers]{natbib}
\begin{document}

\renewcommand\appendixname{Supplemental Material}

\title{Detecting charge transfer at defects in 2D materials with electron ptychography}

\author
{Christoph Hofer,$^{1}$ Jacob Madsen,$^{2}$ Toma Susi,$^{2}$ Timothy J. Pennycook$^{1}$\\
}
\address{
$^{1}$EMAT, University of Antwerp, Campus Groenenborger, 2020 Antwerp, Belgium\\
$^{2}$University of Vienna, Faculty of Physics, Boltzmanngasse 5, 1090 Vienna, Austria
}

\begin{abstract}
Electronic charge transfer at the atomic scale can reveal fundamental information about chemical bonding, but is far more challenging to directly image than the atomic structure. The charge density is dominated by the atomic nuclei, with bonding causing only a small perturbation. Thus detecting any change due to bonding requires a higher level of sensitivity than imaging structure and the overall charge density. Here we achieve the sensitivity required to detect charge transfer in both pristine and defected monolayer WS\textsubscript{2} using the high dose efficiency of electron ptychography and its ability to correct for lens aberrations.
Excellent agreement is achieved with first-principles image simulations including where thermal diffuse scattering is explicitly modeled via finite-temperature molecular dynamics based on density functional theory. The focused-probe ptychography configuration we use also provides the important ability to concurrently collect the annular dark-field signal, which can be unambiguously interpreted in terms of the atomic structure and chemical identity of the atoms, independently of the charge transfer. Our results demonstrate both the power of ptychographic reconstructions and the importance of quantitatively accurate simulations to aid their interpretation. 
\end{abstract}
\maketitle

\section{Introduction}
Chemical bonding makes a material with collective properties out of an ensemble of independent atoms. As a central tenet of density functional theory (DFT)~\cite{hohenberg1964,Kohn1965}, these properties can be entirely derived from the electron charge density. As such, measuring the charge density is of great interest, especially its variation from neutral independent atoms due to bonding. Electron diffraction and X-ray scattering can detect charge transfer~\cite{COPPENS19771,charge_transfer2,Zuo_2004,Wu1999}, but both probe an average over a large area rather than individual atoms or atomic columns. For imaging atomic structures, electron microscopy has become an indispensable tool, and in recent years, scanning transmission electron microscopy (STEM) the de facto standard due to the simplicity of interpreting its annular dark-field (ADF) atomic-number contrast~\cite{Krivanek2010}. However, ADF is not sensitive to the electron charge distribution, as its contrast results from Rutherford scattering from the nuclear cores. The inelastic scattering processes detected by electron energy-loss spectroscopy (EELS) contain information about the local electronic environment~\cite{EELS1,EELS2}, but require high electron irradiation doses as well as the use of especially demanding first-principles spectrum simulations to interpret its fine structure~\cite{EELS1}.

Phase-contrast imaging methods, on the other hand, are known for being highly dose efficient, and indeed, local atomic-scale detection of charge redistribution due to bonding has been demonstrated in high-resolution transmission electron microscopy (HRTEM) for both single-layer hexagonal boron nitride and doped graphene~\cite{Meyer2011}. However, revealing the charge redistribution in HRTEM required careful use of specific contrast transfer function (CTF) conditions with large defocus values, significantly degrading image resolution. Furthermore, due to the oscillatory CTFs of HRTEM, it is often difficult to interpret even for structural imaging, let alone disentangling charge transfer from that of the atomic structure.
Fortunately, STEM phase-contrast imaging has improved greatly in the past decade. Multiple phase-imaging modes can be performed without the need for aberrations and in parallel with $Z$-contrast ADF, which provides easily interpretable images of structures at the maximum resolution of the microscope. Atomic-resolution charge-density imaging has been performed with differential phase contrast (DPC), scattering center of mass (CoM)~\cite{idpc} and off-axis holography~\cite{Boureau2020}, but these techniques have not been sufficient to detect charge transfer due to bonding.

\begin{figure*}
 \includegraphics[width=0.95\textwidth]{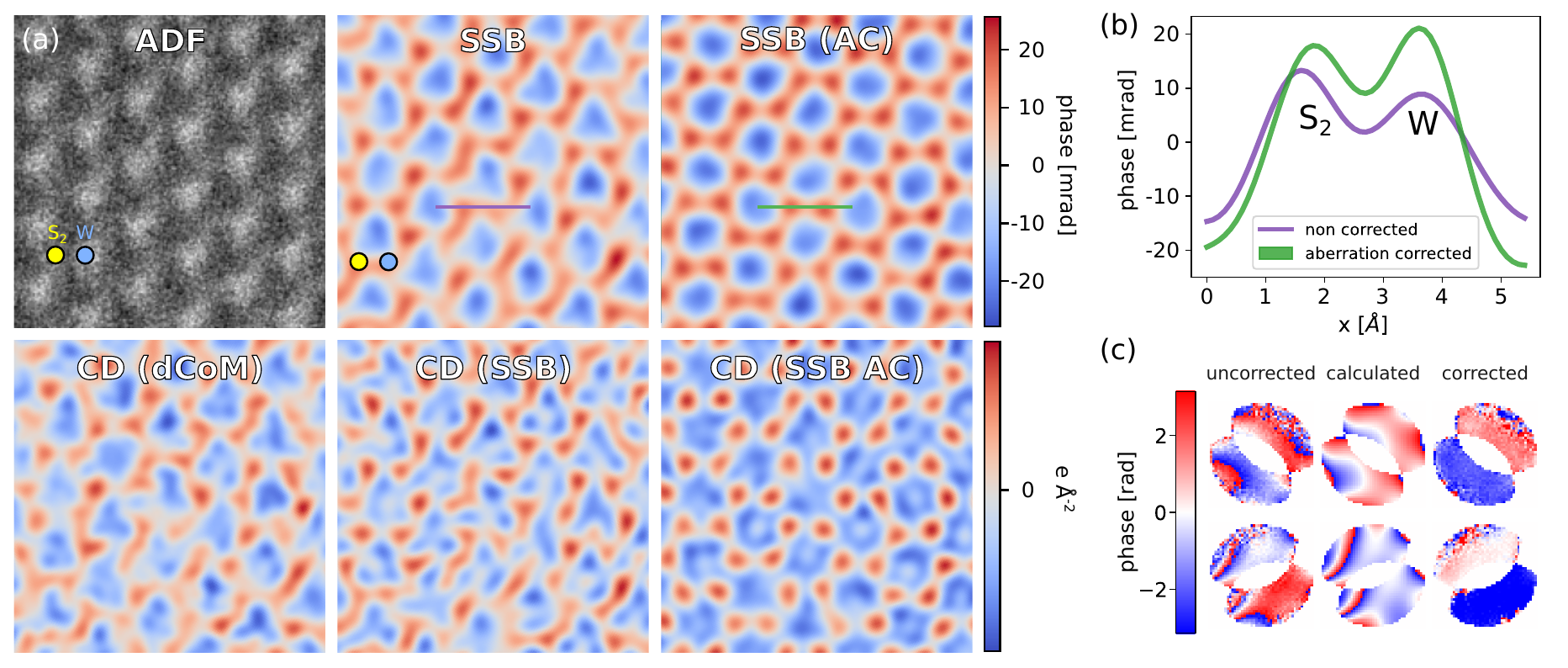}%
\centering\caption{\label{aberrations} \textbf{Residual aberration correction with ptychographic phase and charge density (CD) imaging of WS$_2$.} a) Top row: $Z$-contrast ADF, and simultaneous SSB images before and after  residual aberration correction (AC). Bottom row: The charge density (CD) imaged using conventional differentiated CoM (dCoM) and SSB ptychography before and after residual aberration correction. The dCoM image required Gaussian filtering to make the structure visible above the noise. b) Line profiles from the SSB phase images in (a) showing how residual aberrations can reverse the polarity of the dumbbell, preventing the subtle effect of charge transfer from being detected without the ptychographic aberration correction. c) Phase of the double-disk overlaps at two different spatial frequencies, showing significant variation due to residual aberrations, which is much reduced upon correction.}
 \end{figure*}

Here we use high-speed~\cite{JANNIS20221} focused-probe single-sideband (SSB) electron ptychography with post-collection residual aberration correction~\cite{Yang2016} to provide the necessary precision and accuracy to detect charge transfer directly in WS\textsubscript{2}, including at individual point defects. In addition to the high speed and dose-efficiency of our method allowing us to outrun damage, the ptychographic aberration correction overcomes the problem of residual aberrations that has so far prevented CoM-based methods from achieving valence imaging at atomic resolution~\cite{charge_density,Susana2024}. 
Accounting for residual aberrations, as we do here, is essential because they can completely obscure the effect of charge transfer, and if ignored may result in an entirely incorrect interpretation. Also crucial is the use of our recent parameter-based quantification method~\cite{HOFER2023kernel} that explicitly accounts for the CTF of the imaging modality and overcomes artefacts due to the sample configuration such as mistilt. 

Using the fact that the phase shift is directly proportional to the electric potential for thin samples, analogously to CoM imaging~\cite{LAZIC2016265}, we can take the Laplacian of the SSB phase image to obtain an image that is proportional to the charge density. This allows one to obtain an aberration-corrected charge density map, which is not possible in present CoM or DPC based methods. Here, we show that both ptychographic phase and charge density imaging can reveal the charge transfer, but analysing the phase images is more robust to noise than the charge density map itself because of its reliance on noise-amplifying differentiation~\cite{ROSE1976}.

\section{Results and discussion}
\subsection{Pristine WS\textsubscript{2}}
The problem of residual aberrations is illustrated in Fig.~\ref{aberrations} with experimental data from WS\textsubscript{2} using a modern aberration-corrected STEM (see Methods). Aberrations are particularly problematic when quantifying phase images since even small values can alter the phase~\cite{Lehtinen2015130}. Two SSB images from the same dataset are shown in Fig.~\ref{aberrations}a, one with ptychographic residual aberration correction and one without. Although the data were acquired while using a carefully tuned 5th-order electron-optical aberration corrector, the lattice polarity in the SSB image without the additional ptychographic post-collection correction is reversed because of the residual aberrations, as illustrated by the line profiles in Fig.~\ref{aberrations}b. With such large effects on the phase images resulting from such relatively small aberrations, reliably interpreting uncorrected phase images from either ptychography or CoM-based methods, including DPC, to uncover the subtle effects of charge transfer is in our view essentially untenable.

\begin{figure*}
 \includegraphics[width=0.95\textwidth]{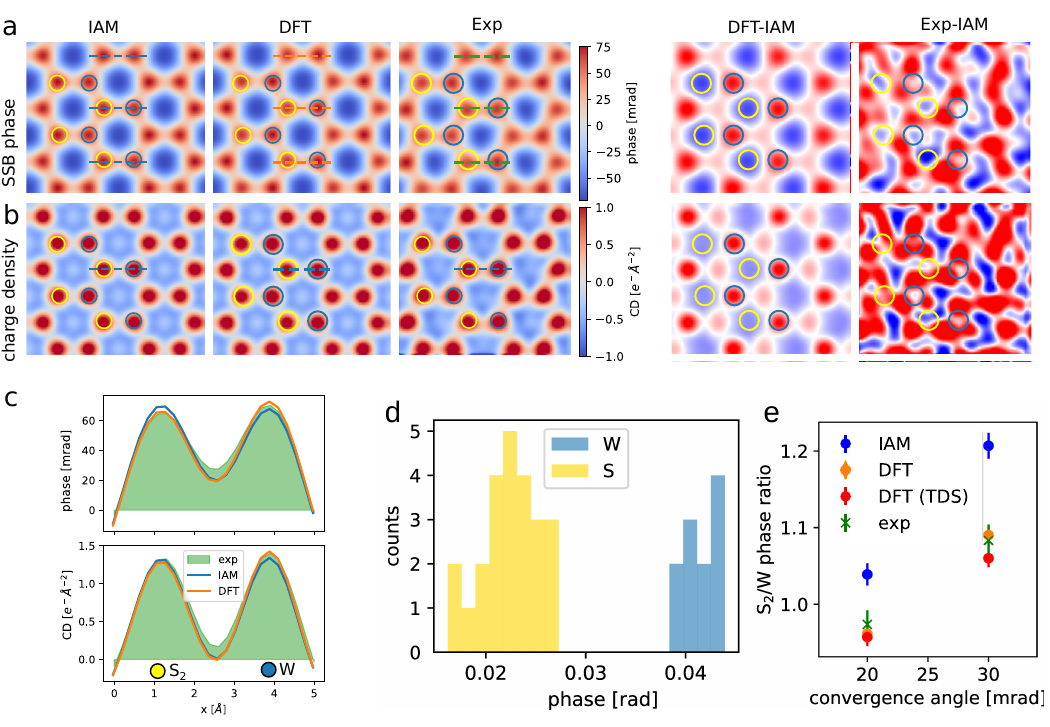}%
 \caption{\label{comparison} \textbf{Detecting charge transfer in WS\textsubscript{2}.} a) SSB phase images reconstructed from simulated and experimental data of pristine WS\textsubscript{2}. The difference between the DFT and IAM phase images results from the electron transfer from W to S. The experimental difference to the IAM-based simulations shows a qualitative match with the theoretical difference. b) Charge-density images based on the SSB phase. The charge transfer is visible in the difference image, but the experimental difference image is noisier than the SSB phase-difference image. c) Line profiles of the experimental and simulated images showing an excellent match between DFT and experiment despite the small difference between the IAM and DFT. d) Histogram of deconvolved S and W phases in the 30~mrad experimental SSB image. e) Ratios of the mean values of the two sublattices using IAM and DFT potentials with and without the effect of thermal diffuse scattering (TDS), compared to experimental data for 20 and 30~mrad convergence angles.
 }
 \end{figure*}

For very thin specimens such as 2D materials, the phase without aberrations should be flat in the double-disk overlaps in probe reciprocal space. The double-disk overlaps from two different spatial frequencies contained in the data are shown in Fig.~\ref{aberrations}c. Without ptychographic aberration correction, the double-disk overlaps of both spatial frequencies are not flat. We use singular value decomposition (SVD) to identify the aberrations present and counteract the phases of these aberrations, resulting in a nearly flat phase in each of the overlap regions, confirming the removal of the residual aberrations~\cite{Yang2016}. See Supplementary Material for additional information.

Fig.~\ref{aberrations}a also shows charge density maps calculated from the CoM and from the ptychographic phase images using differentiation. The raw CoM-based charge density map, the differentiated CoM (dCoM), is very noisy indeed and a Gaussian filter is required to make the structure apparent over the severe noise (see Supplementary Figure~2). The resulting filtered dCoM image shown in Fig.~\ref{aberrations}a is fairly similar to the SSB-based charge-density map without ptychographic aberration correction. The SSB-based charge-density map with the additional ptychographic aberration correction, however, shows a much higher-quality image. As we will show, our post-collection aberration-corrected ptychographic phase images and charge density maps show an excellent match with simulations that include charge transfer.

To understand the role of charge transfer in the observed contrast, we simulated images with and without the effects of bonding. To do so we used potentials based on first-principles DFT and the independent atom model (IAM) as input to multislice 4D-STEM simulations using the \textit{ab}TEM code~\cite{abtem}, with probe parameters and sampling set to match our experimental conditions (see Methods). Fig.~\ref{comparison} shows the ptychographic phase and charge density images simulated based on the DFT and IAM potentials. The figure also shows the differences between the DFT and IAM based images, i.e.~the quantitative difference induced by the bonding. Compared to the IAM the DFT phase image has a higher phase on the W sites and a reduced phase on the S$_2$ sites. Compared to the IAM based ptychographic charge density image, the DFT based result shows a  higher charge density at the W sites and a reduced charge density at the S$_2$ sites.

As the charge redistribution due to bonding contained in the DFT potential is the only significant difference between the DFT and IAM images, the differences between them are attributed to the charge transfer from W to S that occurs with electron redistribution due to bonding (see also SFigs.~3 and 4). Importantly, the magnitude of the difference due to bonding is around 10\% of the maximum values at the atomic sites in the DFT and IAM images, enough to be detected at moderate electron doses (see Supplementary Material). 

Fig.~\ref{comparison}a also shows the result of subtracting the IAM phase and charge density images from the experimental images (scaling the theoretical ones to match the mean and standard deviation of the experimental images). The subtraction is very difficult and prone to error, especially because small mismatches in the alignment of the images will result in large values in the difference images. Despite the difficulty, a qualitative match with the theoretical image is obtained. The experimental difference image does however exhibit a magnitude three times higher than the theoretical difference, especially closer to the center of the hexagons. We expect this difference is mostly due to misalignment caused by experimental scan distortions, sample drift and mistilt, however residual aberrations which are not perfectly accounted for even with post-collection aberration correction may also contribute.

The phase differences between the IAM, DFT and experimental results are also reflected in the line profiles taken across the W-S$_2$ dumbbells shown in Fig.~\ref{comparison}c. An average of 45 dumbbells was used for the experimental profiles to provide high statistical significance. Interestingly, the charge density image provides less statistical significance than the phase image, as the difference image is noisier. 
This is likely because calculating the charge density map involves double differentiation, and differentiation is often associated with noise amplification. We therefore propose that analysing the phase is more reliable when it comes to charge transfer detection.

To quantify the influence of charge redistribution from the ptychographic images, we need to assign a phase to each of the atomic sites. Various approaches have been established for quantifying the atomic intensities of ADF images, including local maxima~\cite{Krivanek2010}, the integration of Voronoi cells~\cite{E2013}, Gaussian fits~\cite{VanAert2009,DeBacker2016}, as well as template matching~\cite{HOFER2021}.
However, these quantification methods do not work well for methods with low transfer of low spatial frequencies such as filtered CoM based imaging and ptychography, as
they induce a nonlinear dependence on the local atomic environment~\cite{HOFER2023kernel}: the proximity of neighbouring atoms can actually decrease the intensity at which they appear in images, which is very different to the additive behavior of probe tails in ADF imaging.  
This introduces a strong image dependence not only on the atomic arrangement but also on sample tilt, which must be accounted for to correctly quantify such phase images, but is not accounted for in any of the usual image quantification methods designed for ADF imaging. 

To overcome this problem, we recently developed a quantification method that specifically accounts for the contrast transfer function (CTF) of the imaging method  and is robust to both sample tilt and source size~\cite{HOFER2023kernel}.
A model fitting the experimental data is initiated and a fast simulation using the convolution between the point-potential of the model and a kernel representing the CTF is optimized with respect to the experimental image. This is especially important here because mistilt is particularly difficult to eliminate in 2D materials and would cause significant errors in quantification, in addition to any uncorrected residual aberrations. With this method, the quantified phase of an atomic column is independent of both the proximity and type of the neighboring atoms. In addition to the SSB used presently, the quantification method can also be used for iterative ptychography and filtered CoM based imaging, which can also exhibit a negative halo around isolated atoms, as we demonstrated in reference~\cite{HOFER2023kernel}.

Fig.~\ref{comparison}d shows a histogram of the quantified phases of the W and S atoms in the experimental image. The phases of the two types of atoms are well separated. The ratio of the mean quantified phase values of S$_2$ to W
depends on the convergence angle as shown in  
Fig.~\ref{comparison}e and, crucially, matches the results of the DFT simulations and not the IAM results. 

The variation with the convergence angle occurs because it determines the strength at which different spatial frequencies are transferred~\cite{YANG2015,OLEARY2021}. This in turn determines the relative strengths at which the features of the charge density distribution appear in the phase images. The excellent agreement of the experiment at both convergence angles with the DFT results, and lack thereof for the IAM results, demonstrates that the frequencies transferred by these conditions contain the information on the charge transfer induced by bonding. We emphasize that such conditions are typical for high-resolution STEM imaging. This is in contrast to the situation in conventional HRTEM, where the imaging conditions relevant to detecting charge transfer are generally not those desired for imaging the structure itself. Including atomic vibrations using thermal diffuse scattering (TDS, see Methods) slightly decreases the S\textsubscript{2} to W ratio compared to neglecting it, and the experimental uncertainty, estimated from the statistical site-by-site phase variation, dominated by the limited signal, overlaps both values with bonding included.

 \begin{figure}
\center\includegraphics[width=0.45\textwidth]{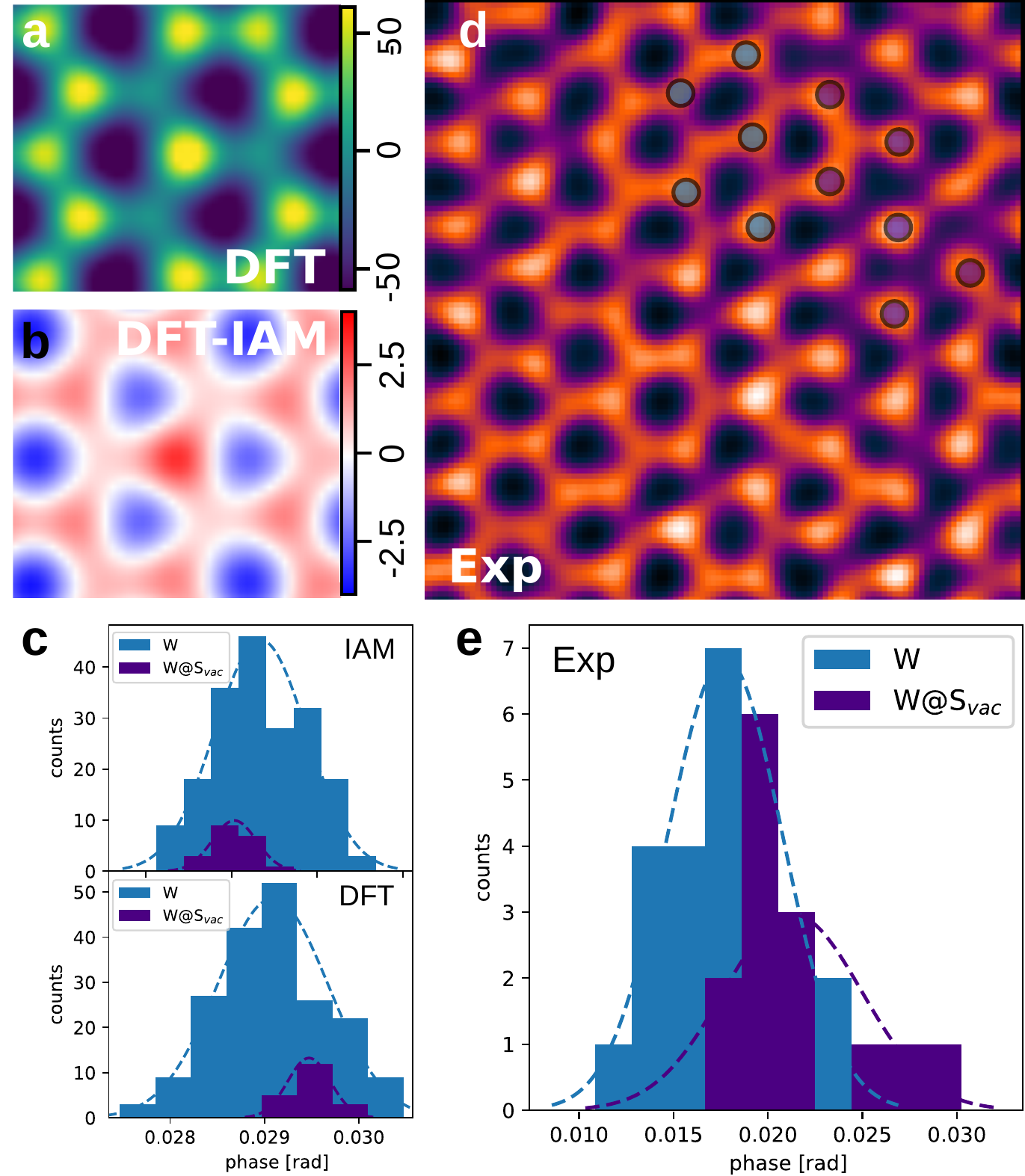}%
 \caption{\label{vacancies} \textbf{Analysis of defective WS\textsubscript{2}.} (a) SSB reconstruction based on a DFT potential of WS\textsubscript{2} with three S vacancies distributed about the central W atom. (b) Difference between IAM- and DFT-based simulations of the same structure. (c) Histogram of extracted deconvolved phases of IAM- and DFT-based simulations of a defective area similar to the experimental defect density, showing a difference between W in the pristine (blue) and the high defect-density region (dark blue) at a $1\times10^5$ e$^-/$\AA$^2$ dose. The statistical distribution is based on 10 randomly noised 4D data sets with this dose. (d) Experimental SSB image of defective WS\textsubscript{2}. The overlaid circles indicate pristine W (blue) and W close to the defects (dark blue), respectively. (e) Histogram of the deconvolved phases of pristine W (blue) and W surrounded by vacancies (dark blue) extracted from the image in (d).}
 \end{figure}
 
\subsection{Defective WS\textsubscript{2}}
The simplest defect in transition metal dichalcogenides is a chalcogen vacancy. For WS\textsubscript{2} such a monovacancy (V$_\mathrm{S}$) can be realized by removing a single S atom, leaving both a S vacancy and one remaining S atom at the former S\textsubscript{2} site. Such vacancies produce a significantly reduced phase at the vacancy sites in SSB reconstructions, as shown in the DFT-based SSB image in Fig.~\ref{vacancies}a, where three S monovacancies (V$_\mathrm{3S}$) are distributed around the central W atom. Note that the S sites are in general not readily visible in the high-angle ADF signal under these conditions, and thus S vacancies are only discernible in the phase images (see also SFig.~7  and 8), highlighting the ability of ptychography to reveal light atoms next to heavy elements~\cite{Yang2016,chuang}.

Interestingly, we observe a higher W site phase close to the S vacancies in our SSB images, with the intensity of the W depending on the density of the vacancies in the area. The reason for this is two-fold: the CTF, and the charge transfer itself. As we discussed earlier, the SSB CTF means that the phase of atoms is reduced in SSB images when they are brought near to each other. Therefore, removing an atom to form a vacancy significantly increases the phase of the neighboring sites in SSB images even for neutral atoms. It is therefore vital to take this CTF effect into account, as we do with our method of phase quantification, allowing us to detect the second effect, the charge transfer.

To systematically study the charge transfer at defects, we simulated images of a region of WS\textsubscript{2} with increasing numbers of S vacancies using both IAM and DFT potentials (SFig.~9). 
For the the V$_{3\mathrm{S}}$ configuration shown in Fig.~\ref{vacancies}a, the difference between the IAM- and DFT-based image simulations is displayed in Fig.~\ref{vacancies}b. Keeping in mind the differences seen in pristine WS\textsubscript{2}, the effect of the vacancies on the difference is greatest at the W atom surrounded by the vacancies. Furthermore, we find that the phase of a W atom in WS\textsubscript{2} increases in the DFT-based images relative to that of the IAM-based images proportionally with the density of S vacancies around the W atom. This is clearly seen in SFig.~9, with the DFT phase at the W atom increasing significantly more with increasing vacancy density than the IAM.
 
\begin{figure*}
 \center\includegraphics[width=0.9\textwidth]{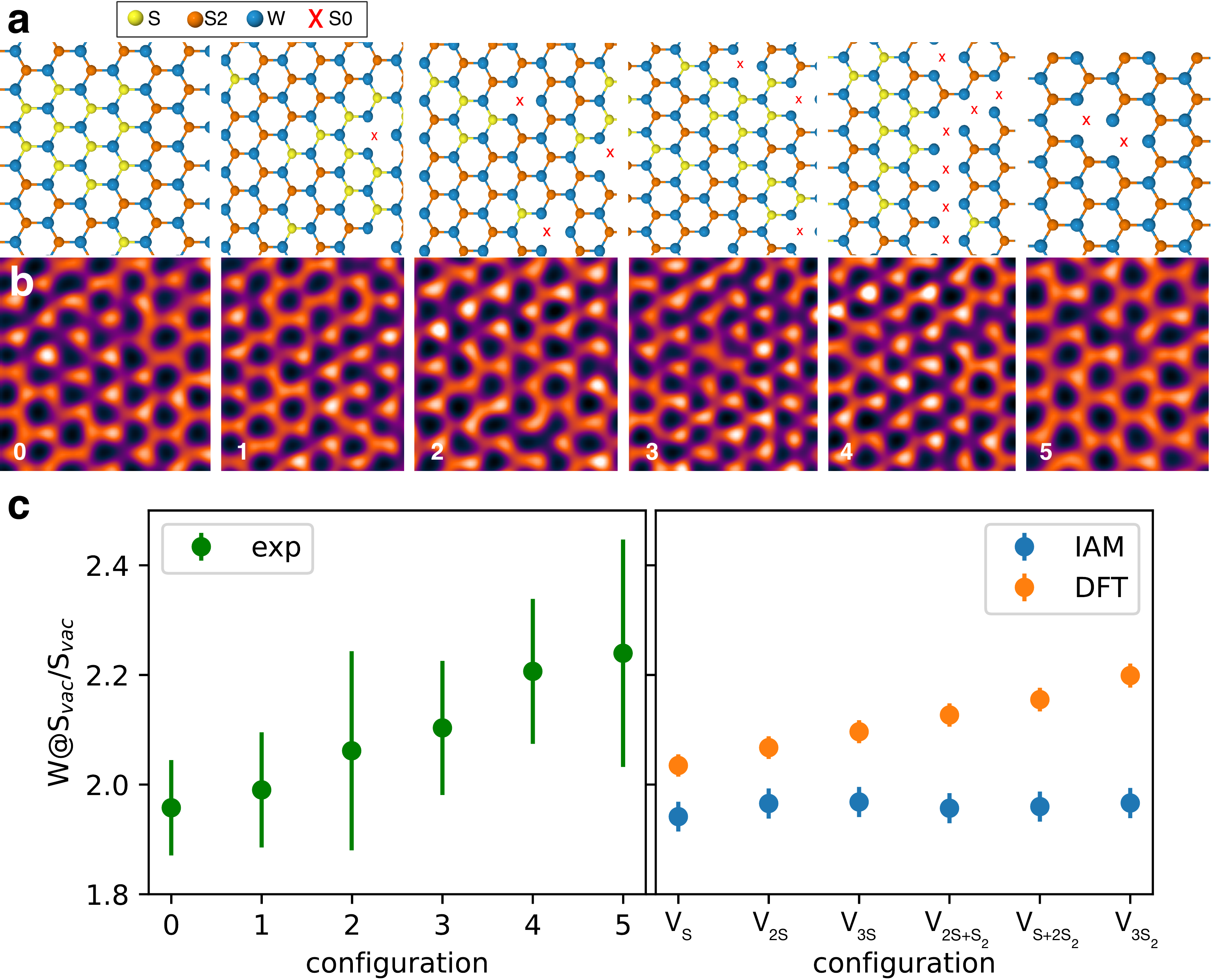}%
 \caption{\label{defect_densities} \textbf{Analysis of the dependence of the $\mathrm{W}@\mathrm{S}_\mathrm{vac}/\mathrm{S}_\mathrm{vac}$ phase ratio on defect density.} (a,b) Atomic models and SSB images of defective WS\textsubscript{2} containing different numbers of mono- and divacancies. (c) Phase ratio between W atoms next to vacancies and at S monovacancies. The variation of ratios clearly depend on the configuration type corresponding to different defect densities. Right: simulated ratios for an increasing density of defects.}
 \end{figure*}

The variation of the quantified phase ratio of the central W site to a S vacancy site ($\mathrm{W}@\mathrm{S}_\mathrm{vac}/\mathrm{S}_\mathrm{vac}$) with respect to the density of vacancies for the DFT- and IAM-based simulations is shown in Fig.~\ref{defect_densities}c. Each subsequent configuration in the simulations has one less S atom. The ratios for the IAM simulations results are essentially flat, while the DFT ratios increase linearly with the density of vacancies. The flatness of the IAM result is exactly what is expected given our quantification method accounts for changes that are purely structural. As the main difference between the IAM and DFT images is the charge transfer in the DFT-based images, the DFT results show that the amount of charge transfer at a W site increases with the amount vacancies surrounding it and that this results in a higher quantified phase of the site.

This increase in the charge transfer with vacancy density would be surprising if one assumes that vacancies reduce charge transfer due to the W atoms having fewer S neighbors. If that were the case, one would see a reduced phase in the DFT images relative to the the IAM images as the vacancy density is increased, rather than the  observed increase. 
In the extreme three-divacancy configuration (V$_\mathrm{3S_2}$), the W atom has lost all its nearest neighbors and one might think is therefore more similar to an independent atom, and expect a smaller difference between the DFT and IAM results. However, this is clearly not the case, as seen in the simulated SSB images and projected all-electron potentials shown in SFigs.~9 and 10 and this highly defected case is in fact the furthest from the IAM result with a higher $\mathrm{W}@\mathrm{S}_\mathrm{vac}/\mathrm{S}_\mathrm{vac}$ ratio than all the other results using the DFT potentials. 
The fact that S vacancies produce occupied states at the defect sites~\cite{Schuler2019,Yuan2014} does not explain the observed contrast, and additional studies have shown charge redistribution occurs towards the defect sites~\cite{SALEHI2016}.

To understand this physical phenomenon, we plot the three-dimensional (pseudo-)electron valence densities of each configuration in SFig.~11, from which a different picture emerges. The line profiles from  W sites adjacent to the defects show a lower density of valence electrons when increasing the number of vacancies around them, in agreement with the increase of phase we observe at this site. The electrons are redistributed towards the next W sites, which creates W-W bonding that is stronger the more S vacancies are involved in the system. Thus we need to consider longer-range charge rearrangement than merely the first-nearest neighbors to understand the full picture.

The relative difference between IAM and DFT phases can reach up to 10\% with vacancies (cf. Fig.~\ref{defect_densities}c). This is a larger difference than in the pristine case, and one might thus conclude that detecting charge transfer is easier at defects. However, defect configurations are often unique and complex and therefore a statistical assessment analysing multiple sites is very challenging with each atom potentially undergoing different amounts of charge transfer. Further, the dose one can use to probe defective configurations is usually lower due to their greater propensity to damage, increasing the uncertainty per site. In this case the dose we could use before damage was only $3\times10^4$ e$^-/$\AA$^2$.

To ascertain whether we can detect charge transfer at defects experimentally, we used electron irradiation to obtain a variety of vacancy densities and atomic configurations in the sample. A representative area is shown in Fig.~\ref{vacancies}d, with the full field of view shown in SFig.~7. 
Fig.~\ref{vacancies}c shows histograms of the extracted phases of the W atoms in the pristine area (blue) and the W atoms adjacent to vacancies, labeled as $\mathrm{W}@\mathrm{S}_\mathrm{vac}$ (dark purple) for the IAM- and DFT-based simulations. Both types of simulations were treated with Poisson noise in the 4D diffraction data corresponding to the experimental dose. The calculated mean value for $\mathrm{W}@\mathrm{S}_\mathrm{vac}$ shows a significant increase in phase compared to pristine W atoms in the DFT simulations. Note that the structure analyzed here only contains monovacancies (cf. SFig.~8a,b); and that the phase shift due to charge transfer would be significantly higher if this specific area of interest also contained divacancies.
This agrees with the experimental data, where an increase in the mean phase of the  $\mathrm{W}@\mathrm{S}_\mathrm{vac}$ compared to that of the pristine W sites is indeed observable in Fig.~\ref{vacancies}e.  

To further study the effect of the vacancies on the charge transfer experimentally, we calculated the phase ratio of the sites between which the majority of the charge transfer occurs,  $\mathrm{W}@\mathrm{S}_\mathrm{vac}/\mathrm{S}_\mathrm{vac}$, using the five different regions of the sample shown in SSB images in Fig.~\ref{defect_densities}b. Each region has a different vacancy configuration, labeled from zero to five, and a model for each is shown in Fig.~\ref{defect_densities}a. The experimental ratios are shown in Fig.~\ref{defect_densities}c alongside the values calculated from the DFT- and IAM-based simulations. Both experimental and DFT configurations are presented in their respective plots in order of increasing vacancy density around the measured $\mathrm{W}@\mathrm{S}_\mathrm{vac}$. Of the experimental configurations, zero has the lowest average number of vacancies around the $\mathrm{W}@\mathrm{S}_\mathrm{vac}$ and five has the highest. 

The low doses used in the experiment produce significant error bars particularly for the experimental configurations with very few $\mathrm{W}@\mathrm{S}_\mathrm{vac}$ such as configuration five with only five $\mathrm{W}@\mathrm{S}_\mathrm{vac}$. However the overall effect seems clear: as the density of vacancies around the $\mathrm{W}@\mathrm{S}_\mathrm{vac}$ increases so too does the ratio $\mathrm{W}@\mathrm{S}_\mathrm{vac}/\mathrm{S}_\mathrm{vac}$, entirely consistent with the findings of our theoretical calculations.  
Experimental configuration five contains only double vacancies, and they are next to each other, meaning the $\mathrm{W}@\mathrm{S}_\mathrm{vac}$ are either next to one or two double vacancies. The calculated ratio
for this configuration is the highest, consistent with those of the DFT configurations containing divacancies. Experimental configuration zero, on the other hand, contains no divacancies and has a much lower ratio, more consistent with the DFT calculations containing only monovacancies. 

Clearly we are hindered in our ability to quantify the charge transfer itself in our experiments at the level of precision imposed by the dose required here to avoid damaging the defects, as well as the small sample sizes used. Furthermore the experimental structures are generally part of a larger, more complex structural environment than captured by our DFT calculations. Nonetheless, the increase in the mean experimental ratios with the density of vacancies at the measured sites fits very well with those predicted from theory. It therefore seems clear we can indeed detect the charge transfer caused by defects and indeed are sensitive to their density.

\section{Conclusions}
To conclude, we have experimentally detected charge transfer in monolayer WS\textsubscript{2} including at its defects, via the influence of the electron distribution in the material on the phase of the probe electrons using electron ptychography and a parameter-based quantification method. Post-collection aberration correction is crucial to obtain accurate phases for charge-transfer measurements, and the convergence angle is also an important parameter, as it defines which spatial frequencies are transferred. Our results indicate that the frequencies at which charge transfer is detectable conveniently include those used for atomic-resolution imaging. Simulations based on first-principles charge redistribution due to bonding lead to an excellent agreement in the pristine case. For defective areas, we observe a notable phase increase of the metal site close to chalcogen vacancy sites, which significantly increases with the number of vacancies involved, as shown both theoretically and experimentally. Our study thus presents a significant advance in detecting atomic-scale charge transfer at defects.

\section{Methods}
\subsection{Sample preparation}
WS\textsubscript{2} was synthesized on Si/SiO\textsubscript{2} by physical vapor deposition from WS\textsubscript{2} powder. The flakes were then transferred to a Quantifoil(R) TEM grid by a drop of isopropanol. The transfer was completed by etching the SiO\textsubscript{2} with a KOH solution and cleaning by deionized water.

\subsection{Transmission electron microscopy}
STEM experiments were conducted on a ThermoFisher Themis Z instrument equipped with a Timepix3 camera at 60~keV. The probe convergence angle was set to 30 mrad unless noted otherwise, and the beam current was ca. 2~pA. Typical 4D datasets included 1024$\times$1024 probe positions with a dwell time of 1--5~$\mu$s. The list of events provided by the Timepix3 was converted to 4D data using an in-house code. 
The detector sampling was chosen to be approx.\ 30--40 pixels for the bright field disk and the real-space sampling was ca.\ 0.1--0.15~\AA\ per pixel. The ADF detector semi-angular range was 70 to 220~mrad. To increase the signal-to-noise ratio, several frames (up to 10) where aligned and averaged. Typical resulting total doses were approx.\ $10^5$~$e^-/$\AA$^2$.

\subsection{Density functional theory}
For DFT, we used the projector-augmented wave method in the open-source package GPAW~\cite{gpaw}. The hexagonal unit cell of WS$_2$ was made orthogonal, its size was optimized, and then used to create 5$\times$3$\times$1 supercells of WS$_2$ (a total of 90 atoms) with 8.9~\AA\ of vacuum between the periodic images of the layers (perpendicular cell size of 12~\AA).
To obtain relaxed structures for the multislice simulations, the atomic positions were optimized using a plane-wave basis with a cutoff energy of 500~eV and 3$\times$3$\times$1 \textbf{k}-points to sample the Brillouin zone until residual forces were below 0.02~eV/\AA. To create models with defects, we removed one or more S atoms, and re-relaxed the atomic positions.

To verify the convergence of the computational parameters, we further relaxed selected models (pristine, 3MV, 3DV) using a higher plane-wave cutoff of 800~eV, a denser \textbf{k}-point mesh of 5$\times$5$\times$1, and included spin polarization. For these DFT potentials, we ran the full 4D-STEM simulations followed by the single-sideband reconstruction and kernel optimization (see below). Differences in the phase ratios of the sites with respect to the default parameters were at most 0.5\%, which we have include in the uncertainty estimates for our computational values.

Charge transfer between the W and S$_2$ columns was estimated via Bader partitioning~\cite{bader} of the all-electron charge density using the grid-based implementation of Henkelman and colleagues~\cite{tang_grid-bader}. The resulting atomic charges for the pristine system were compared to those calculated for atoms surrounding the S vacancy.

\subsection{Multislice simulations}
For the multislice simulations, the DFT potential was calculated from the all-electron charge density converged with GPAW, as described previously~\cite{susi_scattering_factors}. For the IAM, the potential was taken to be a superposition of individual tabulated atomic potentials, as parametrized by Lobato~\cite{lobato}. A lateral real-space sampling of 0.05~\AA\ has been used for both potentials as well as a perpendicular slice thickness of 1.0~\AA.

Thermal diffuse scattering was included by running DFT-based velocity-Verlet molecular dynamics (MD) with a timestep of 2~fs, initialized with a Maxwell-Boltzmann temperature of 300~K and thermalized for 100 timesteps. Ten MD frozen-phonon DFT-potential snapshots were generated by saving the converged electron density every 50 timesteps, and datasets averaged over the ensemble. Although more images would likely be required for full convergence of an atomically thin material, the cost of DFT/MD for large supercells is high and considering the small effect of TDS on the phase ratios, this does not affects our conclusions.

4D-STEM simulations were carried out using the multislice approach as implemented in the open-source package \textit{ab}TEM~\cite{abtem}. All simulation parameters were set to the experimental parameters, including the convergence angles and the electron dose per area simulated as Poisson noise of the diffraction patterns. To estimate statistical variation caused by the limited dose, multiple randomly noised datasets were analyzed as indicated in the main text. The reciprocal-space sampling was 0.06~\AA$^{-1}$. For ADF, we note that DFT and IAM potentials produce essentially identical contrast.

\subsection{Single-sideband ptychography}
SSB ptychography was performed with the open-source PyPtychoSTEM package~\cite{gitlab}, using either the experimental or simulated 4D datasets as input. The step size was 0.1--0.15~\AA\ per pixel depending on the dataset and the voltage was 60~kV. The convergence angle was set to 20 or 30~mrad. Post-collection aberration correction was applied using singular value decomposition to identify the residual aberrations which were then counteracted. More details are given in the Supplementary Material.    

\subsection{Kernel method for assigning phases to atomic sites}
The phases were extracted by an optimization method where atomic phases are fitted to the target image~\cite{HOFER2023kernel}. First, an initial atomic model that represents the experimental configuration is aligned with the target image. The model is converted to a point potential, which is convoluted with a kernel calculated by the contrast transfer function of SSB. The resulting simulation is compared with the target image and their correlation is maximized by optimizing the alignment between the model and the experimental image, the width of the kernel and the intensities of the point potential. For the experimental data, the sample tilt is also optimized. Correlations were above 99\% for the simulated data and above 93\% for the experimental data.

\vspace{12pt}
\section*{Acknowledgement}
We acknowledge funding from the European Research Council (ERC) under the European Union's Horizon 2020 Research and Innovation Programme via Grant Agreement No. 802123-HDEM (C.H. and T.J.P.) and No.~756277-ATMEN (J.M. and T.S), and FWO Project G013122N “Advancing 4D STEM for atomic scale structure property correlation in 2D materials” (C.H.). We further acknowledge computational resources provided by the Vienna Scientific Cluster (VSC).


\bibliography{main}

\appendix

\title{Supplementary Material}
\renewcommand{\figurename}{Supplementary Figure}
\renewcommand{\tablename}{Supplementary Table}
\setcounter{figure}{0}
\setcounter{page}{1}
\renewcommand{\thesubsection}{S\arabic{subsection}}

\subsection{Post-acquisition aberration correction}

\begin{figure*}[h]
 \center\includegraphics[width=1.0\textwidth]{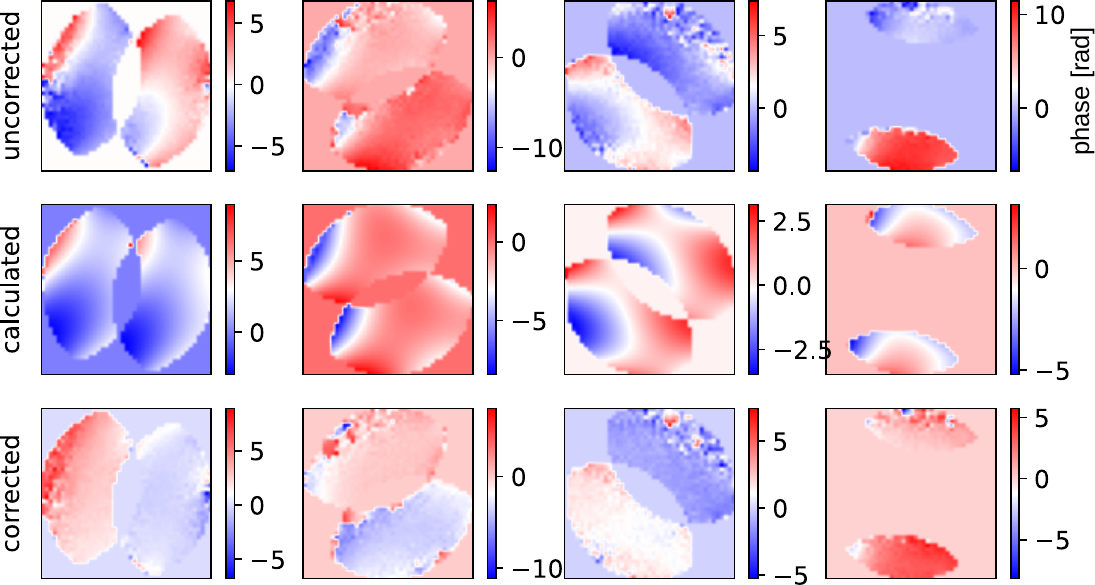}
 \caption{\label{trotters} \textbf{Unwrapped phases of the double-disk overlaps (DDOs) at different frequencies.} Top: Uncorrected DDO phases. Middle: SVD-calculated DDO phases used for the aberration correction. Bottom: Compensated DDO phases.}
 \end{figure*}

Correcting the residual aberrations in phase images is crucial to determine the phase changes imposed by the sample accurately. In SSB, the double-disk overlaps in probe reciprocal space can be used to analyze any aberrations present. SFig.~\ref{trotters} shows examples of the double-disk overlaps from one of our datasets for four different spatial frequencies. The top row shows the experimental phases, the middle row the calculated phases with the identified aberrations and the bottom row the compensated phases, which are much more flat. The phase of the double-disk overlaps should be flat in the absence of aberrations. The residual aberrations identified by singular value decomposition (SVD) in the data shown in Fig.~\ref{trotters} are listed in Table \ref{aberrcoeffs}. Note that we used a fifth order electron optical aberration corrector, which makes the magnitudes of the lower orders higher to compensate the higher-order aberrations.

\begin{table}
\caption{Aberration coefficients up to 5th order of the experimental data shown in Fig.~1 of the main text.}
\begin{center}
\begin{tabular}{| c c |}
\hline
C10  &  3.923  nm\\
C12a &  4.494  nm\\
C12b &  0.719  nm\\
C21a &  425.722  nm\\
C21b &  -538.706  nm\\
C23a &  -29.749  nm\\
C23b &  -173.962  nm\\
C30  &  -18.358  $\mu$m\\
C32a &  -23.239  $\mu$m\\
C32b &  -30.227  $\mu$m\\
C34a &  2.782  $\mu$m\\
C34b &  -0.213  $\mu$m\\
C41a &  -0.491  mm\\
C41b &  0.322  mm\\
C43a &  -0.292  mm\\
C43b &  0.369  mm\\
C45a &  -0.083  mm\\
C45b &  -0.134  mm\\
C50  &  7.208  mm\\
C52a &  24.914  mm\\
C52b &  34.317  mm\\
C54a &  -9.11  mm\\
C54b &  10.464  mm\\
C56a &  -5.105  mm\\
C56b &  8.305  mm\\[1ex] 
\hline
\end{tabular}
\end{center}\label{aberrcoeffs}
\end{table}

\subsection{dCoM based charge density maps} 
\begin{figure*}[h]
 \includegraphics[width=\textwidth]{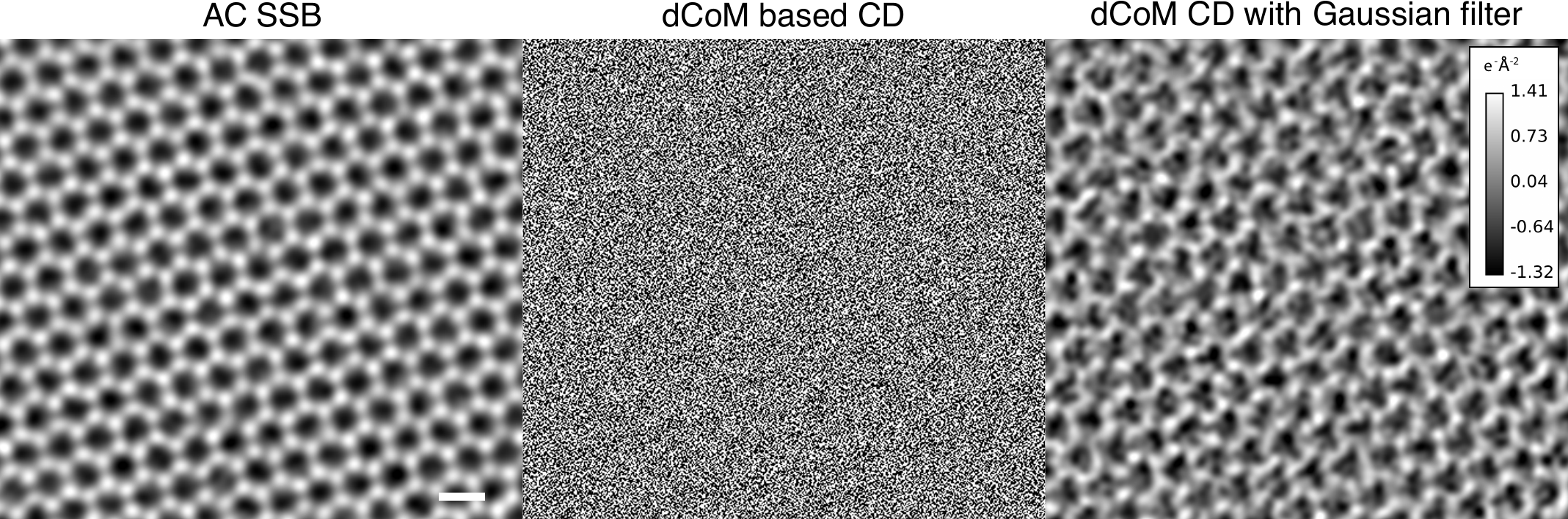}
 \centering\caption{\label{CD} \textbf{SSB phase image and dCoM based charge-density (CD) maps of pristine WS\textsubscript{2}.} The dose is approximately $5\times10^4$ e$^-$\AA$^{-2}$. The left image shows the SSB image with ptychographic aberration-correction (AC). The middle image is the raw dCoM based charge-density map, which is very noisy due to the limited amount of electrons. Gaussian filtering reveals the atomic structure (right).}
 \end{figure*}

SFig.~\ref{CD} illustrates the benefits of ptychographic phase imaging over direct charge-density imaging based on the CoM data. The dCoM charge-density map is not only affected by aberrations, but is also far noisier at a given dose. A Gaussian filter reveals the atomic structure, but the strong variation over the atomic sites in the filtered images  is still clearly too high to quantify the charge transfer. The ptychographic phase image with post-collection aberration correction  on the other hand is far clearer and sufficiently sensitive to detect the charge transfer.

\begin{figure*}[!h]
 \includegraphics[width=\textwidth]{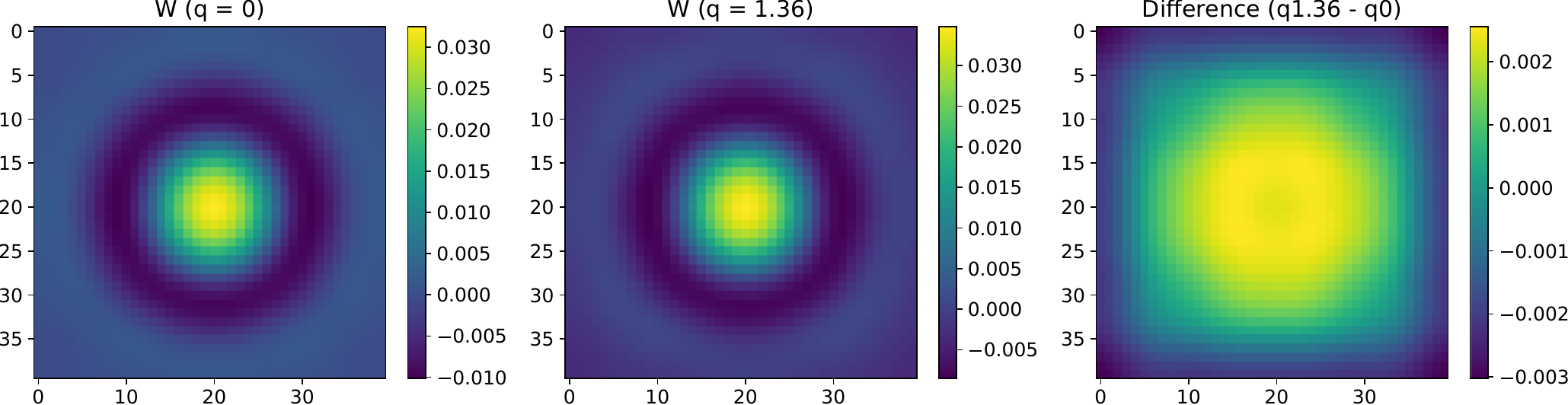}
\centering \caption{\label{SSB_W} \textbf{Phase change induced by charge removal.} Left: Phase image of an independent neutral W atom. Middle: Phase image of a W atom with a removed charge of 1.36 electrons. Right: Difference between the two.}
 \end{figure*}

 \begin{figure*}[!h]
 \includegraphics[width=\textwidth]{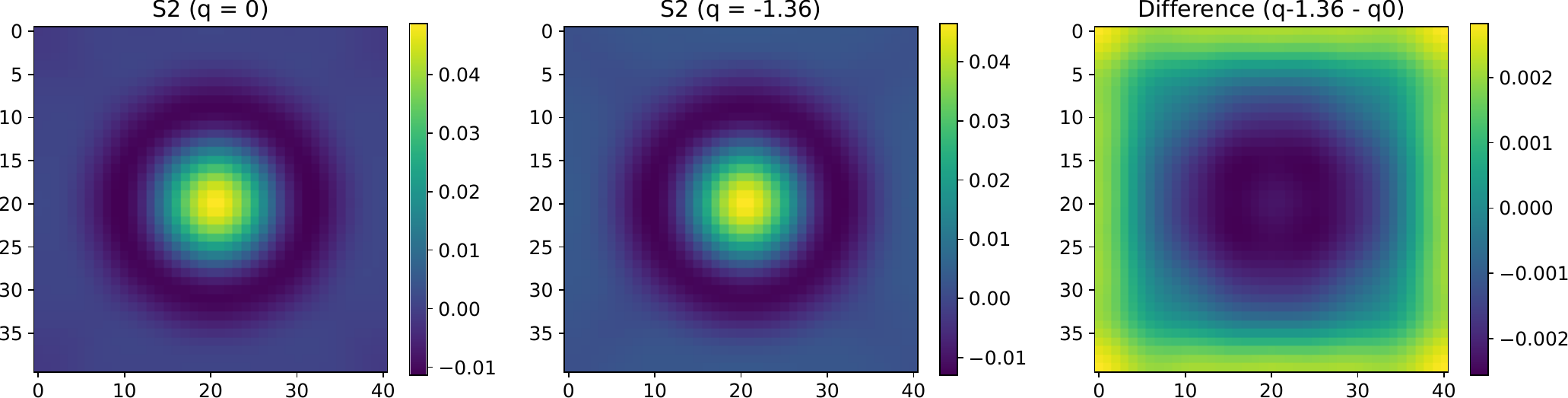}
 \centering\caption{\label{SSB_S2} \textbf{Phase change induced by charge addition.} Left: Phase image of an isolated neutral S\textsubscript{2} column. Middle: Phase image of a S\textsubscript{2} column with an added charge of 1.36 electrons. Right: Difference between the two.}
 \end{figure*}
\clearpage

\subsection{Effect of charge transfer on phase shifts}
In pristine WS\textsubscript{2}, the less electronegative W atoms donate 1.36 electrons to the neighboring S. Using DFT potentials for W atom SSB ptychography simulations, we can study the effect of charge removal or addition on the SSB phase, as shown in SFigs.~\ref{SSB_W} and \ref{SSB_S2}. The left panels show an isolated W atom or S\textsubscript{2} column, which corresponds to an IAM model. The middle panels show the phase for these models with an additional / reduced charge $q$ equal to 1.36 elementary charges, i.e.\ to a removal / addition of 1.36 electrons, corresponding to the charge transfer in WS\textsubscript{2}. As indicated by the difference between the two plotted in the right columns, the phase change is up to 8\% for the W and -5\% for S\textsubscript{2}. In practice, the differences are smaller as charge is not completely removed into a uniform compensating background charge as required by DFT, but rather transferred to nearby sites. This example corresponds to perfect ionicity of the chemical bond, while in practice any covalency reduces the contrast. The crucial point, however, is that due to its effect on screening the nucleus, charge transfer can result in an order of magnitude larger change in the phase than in the potential itself, which only changes by 0.5\%. This larger change in the phase substantially aids the detection of charge transfer.

\subsection{IAM parametrization}
Another potential source of uncertainty in our analysis of the simulated phase ratios are unavoidable differences in how the IAM and DFT potentials are constructed. From GPAW, we recover the exact core-electron density from the projector augmentation functions, to which the self-consistent valence density is added alongside the nuclear contributions. An IAM such as the Lobato parametrization, on the other hand, is a fit to an atomic radial potential typically obtained from a relativistic Hartree-Fock calculation, with a limited number of fitting parameters and a specific functional form. Although care has been taken in \emph{ab}TEM to ensure that DFT and IAM potentials are constructed in a consistent way, phase differences due to charge transfer are so small that minor numerical details might matter.

\begin{figure}[h]
 \includegraphics[width=0.85\textwidth]{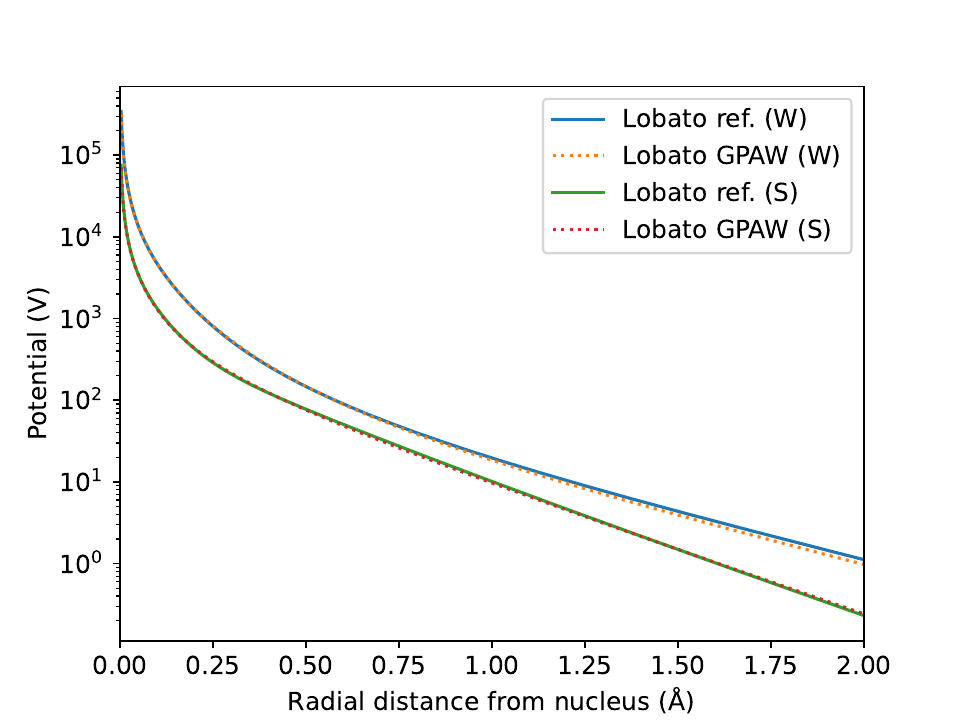}%
 \caption{\label{lobato} \textbf{Comparing IAM parametrizations.} Radial potentials calculated for W and S atoms based on the IAM parametrization of Lobato (solid lines) compared to a custom parametrization of the same functional form fitted to an atomic GPAW calculation (dotted lines). Minor differences can be observed further away from the nuclei (note the logarithmic scale of the vertical axis).}
 \end{figure}

To estimate this effect, we ran GPAW calculations for isolated neutral W and S atoms, and fitted the resulting radial potentials with the parametrization used by Lobato to create a custom IAM based on our DFT potential. SFig.~\ref{lobato} shows the fitted potential compared to the original Lobato parametrization, with only very minor differences further away from the nucleus. We then performed our full analysis (generation of potentials, 4D-STEM simulation, SSB reconstruction, kernel optimization) on isolated W and S\textsubscript{2} columns using both potentials, and found a 0.9\% difference in the S\textsubscript{2}/W phase ratio between the two parametrizations. Although our full DFT potential may arguably be more accurate, we have nonetheless included this as an uncertainty for our simulated IAM values in the main text.

\begin{figure*}
 \includegraphics[width=\textwidth]{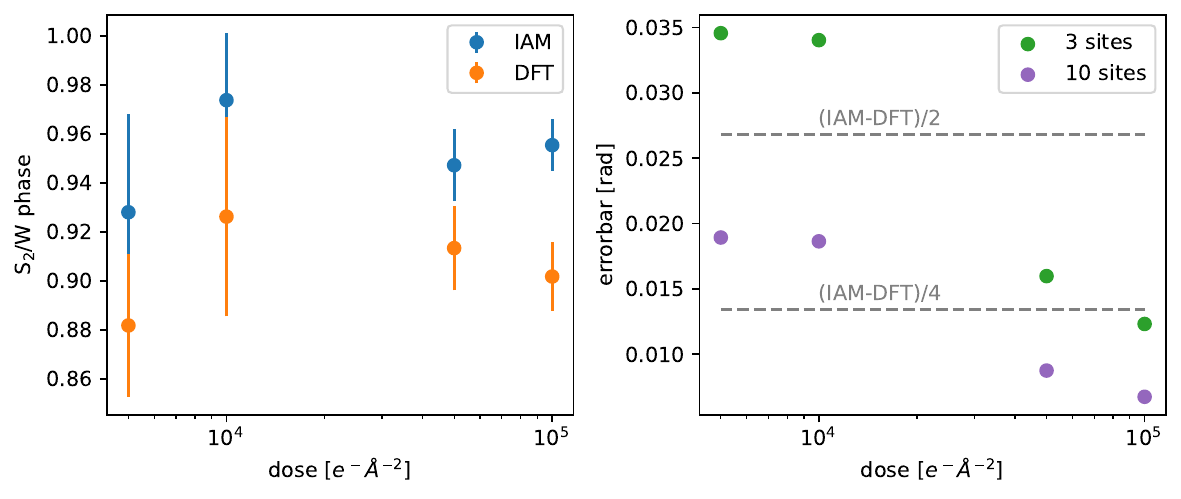}%
 \centering\caption{\label{dose} \textbf{Accuracy of phase extraction.} Left: Ratio between the S\textsubscript{2} and W site phases at different doses. The error bar is calculated by the variation of phases as a function of Poisson noise. Right: Comparison of the ratios with different numbers of sampled sites. The horizontal dashed lines denote limits precision as fractions of the difference between IAM and DFT simulations.}
 \end{figure*}
 
\subsection{Phase uncertainty due to limited dose}
Experimentally, sufficient precision is either achieved with high doses, combining measurements from multiple identical atomic sites, or both. Dose is the most limiting factor when it comes to phase-difference measurements, thus averaging over multiple sites is often unavoidable. The amount of electron dose a specimen can handle until damage occurs depends very much on the material and typical doses for imaging 2D materials range from 10$^4$--10$^6$ $e^-$\AA$^{-2}$. The difference between the S\textsubscript{2}/W ratios have to be sufficiently separated so that their uncertainties are not overlapping. The error bars, which are calculated by the variation of phases due to noise, depend on the dose as well as on the number of atoms analysed. 
Thus obtaining sufficient statistical precision for configurations with small numbers of atoms is much more difficult. 

To estimate the dose-dependent precision of the phase extraction of charge transfer due to bonding for WS\textsubscript{2}, we show in SFig.~\ref{dose} the effect on the quantified S\textsubscript{2}/W phase ratio for the IAM and DFT potentials using first just three pairs of sites. Since a ratio is calculated in this analysis, the relative errors between both atomic sites are summed for the total relative error. The uncertainties are sufficiently small to distinguish the IAM (no bonding) and DFT (bonding) simulations for doses of $5\times10^4$ $e^-$\AA$^{-2}$ and above for three measurements. The standard error of the mean decreases with the square root of the number of sites, $\sqrt{N}$. Thus for instance with 10 pairs of sites, the influence of charge transfer is significant compared to the uncertainty at doses at least as low as $5\times10^3$ $e^-$\AA$^{-2}$.
 
\subsection{Phase shifts at defective sites}
\begin{figure*}[b]
 \includegraphics[width=1\textwidth]{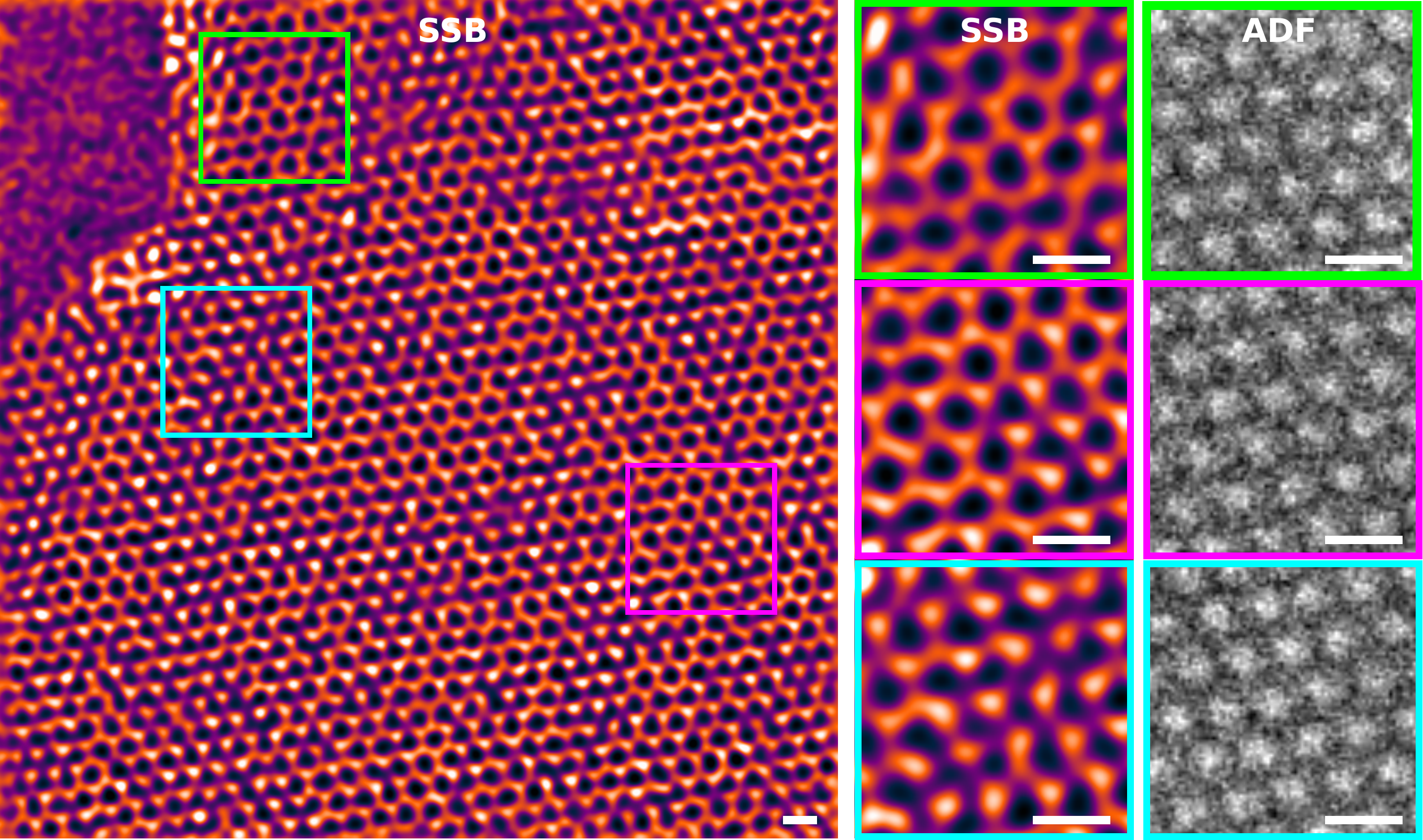}%
 \centering\caption{\label{areas} \textbf{Experimental data from a defected region of WS$_2$. Regions of different vacancy configurations and defect densities are highlighted on the right.} The green box shows a pristine region, the purple box mono vacancies and the blue box a region with divacancies. These atomic structure of the defect sites can be only identified in the SSB images.}
 \end{figure*}

SFig.~\ref{areas} shows a large area of the defective WS\textsubscript{2}, with smaller regions highlighted on the right side. The green box shows a small region of pristine lattice, the purple box a region with several S vacancies, and the cyan box a full line defect. In all cases, the SSB shows a much clearer image of the defects than the ADF image; it is difficult to even identify the S sites in the ADF signal. Besides this, the phase of W is also higher at sites with a higher defect density. This is due to both the contrast mechanism of SSB as well as charge transfer. All six different configurations discussed in the main text are present in this image.

\begin{figure*}[b]
 \includegraphics[width=1\textwidth]{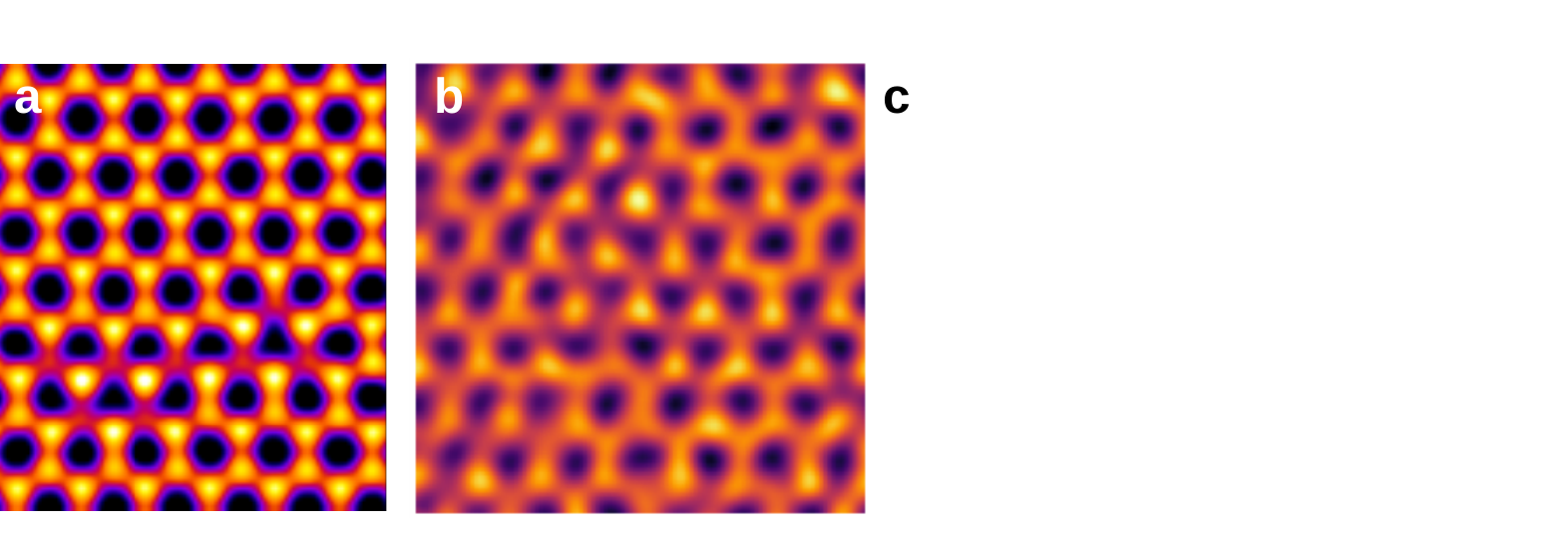}%
 \centering\caption{\label{def} \textbf{Analysis of phase cross sections at a defective area.} a) Simulated area that is analysed for Fig.~\ref{vacancies}c in the main manuscript. b) Experimental SSB image of a small defective area. c) Extracted phases of (b).}
 \end{figure*}

SFig.~\ref{def}a shows a simulation of the defective area that is used for the analysis in Fig.~\ref{vacancies} of the main text containing several monovacancies. SFig.~\ref{def}b shows another example of a small region of the experimental SSB image of a defective region. The experimental phases from all the sites in SFig.~\ref{def}b are extracted and shown in the corresponding histograms on the right. For this specific region, the S vacancies, the pristine sites (S$_2$ and W) and the W close to the S vacancies (W@S\textsubscript{vac}) are identified manually. As one can see, the W@S\textsubscript{vac} sites have a higher phase than the corresponding pristine W. This is a result of the charge transfer at the defective sites, as discussed in the main text. The magnitude of the shift is related to the amount of charge transfer.
 
\begin{figure*}
 \includegraphics[width=0.9\textwidth]{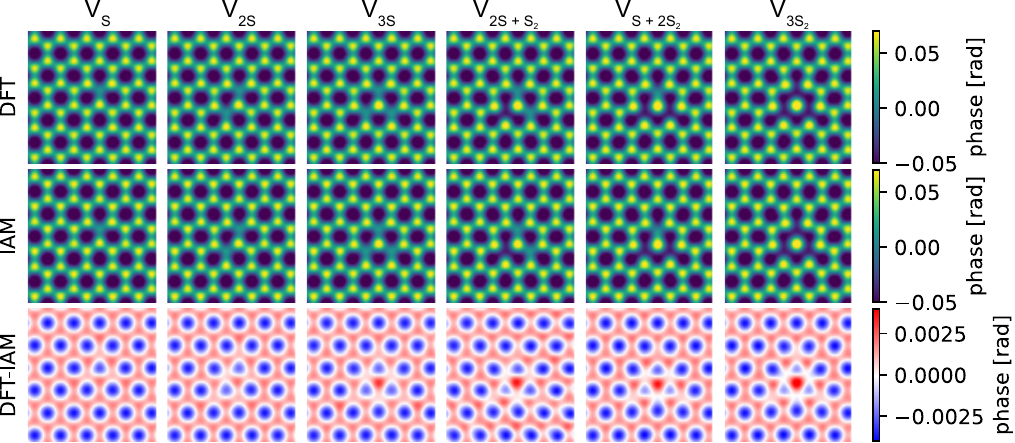}%
 \caption{\label{diff1} \textbf{SSB ptychography images simulated based on DFT and IAM potentials and their difference for different vacancy configurations.} As the density of vacancies increases (left to right) more charge is transferred, resulting in greater contrast in the DFT--IAM image. The configurations correspond to a single monovacancy (MV), two MVs, three MVs, a single divacancy (DV), two DVs and three DVs.}
 \end{figure*}

SFig.~\ref{diff1} shows SSB image  simulations of structures with an increasing number of vacancies inserted around a W atom. Simulations based on both the DFT and IAM potentials as well as the difference between the corresponding SSB images are presented. The difference images show the result of the charge transfer, as the charge transfer is the only significant difference between the DFT and IAM images of a given model. The first three columns show one to three monovacancies (MVs) surrounding a W site. In the last three columns divacancies (DVs) are formed by removing the remaining S atoms from the MV sites, meaning there is a systematic increase of the density of vacancies from left to right. The image contrast is kept constant, with a set range of displayed phase values for each row of images in the figure. It is clear that the difference between the DFT and IAM increases with the vacancy density, especially at the W site around which the vacancies are introduced. 

 \begin{figure*}[b]
 \includegraphics[width=0.9\textwidth]{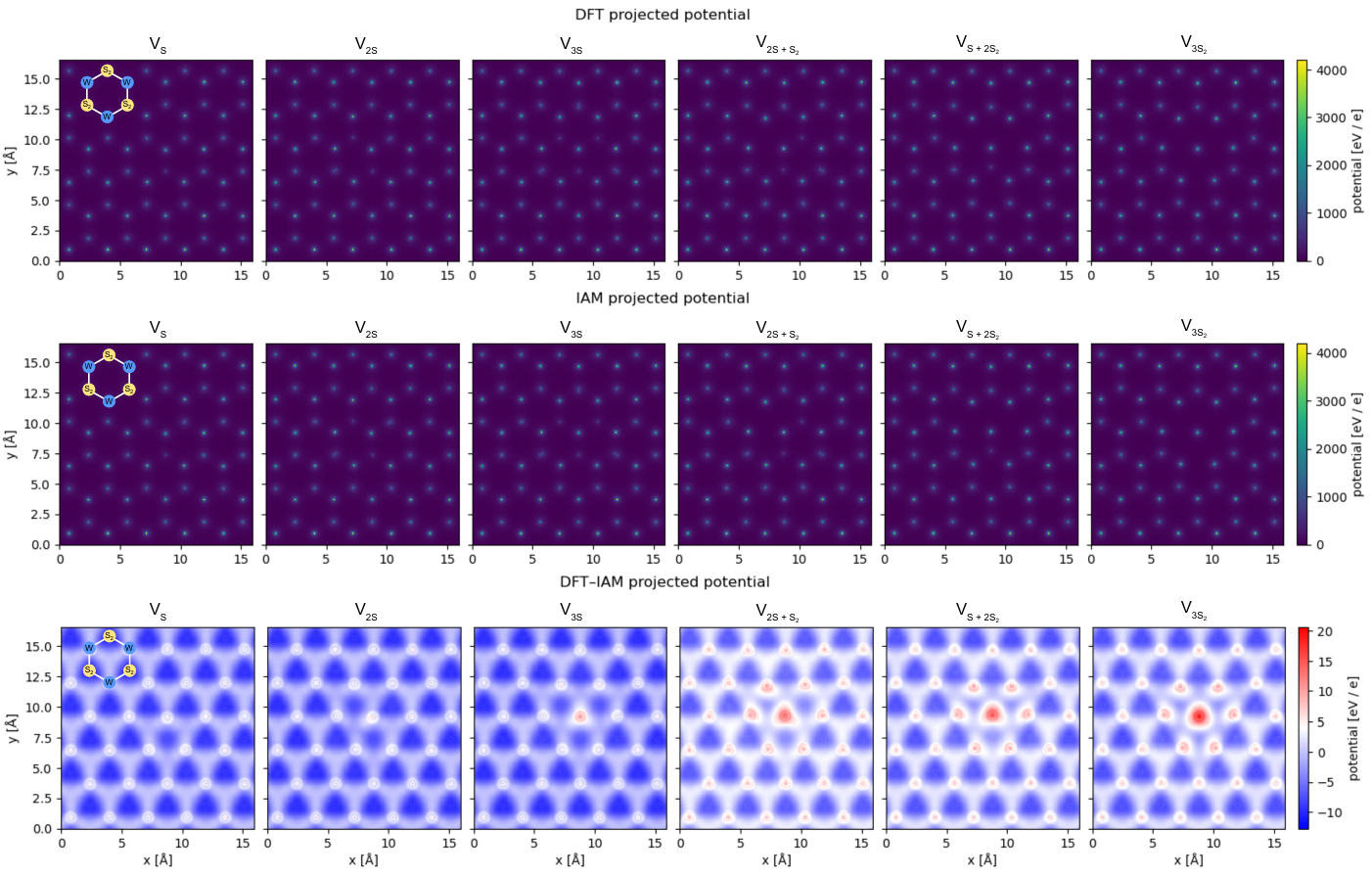}%
 \centering\caption{\label{diff} \textbf{Projected DFT and IAM potentials and their difference for the vacancy configurations shown in SFig.~\ref{diff1}.} The density of defects (vacancies) increases left to right.}
 \end{figure*}
 \begin{figure*}[h!]
 \includegraphics[width=0.44\textwidth]{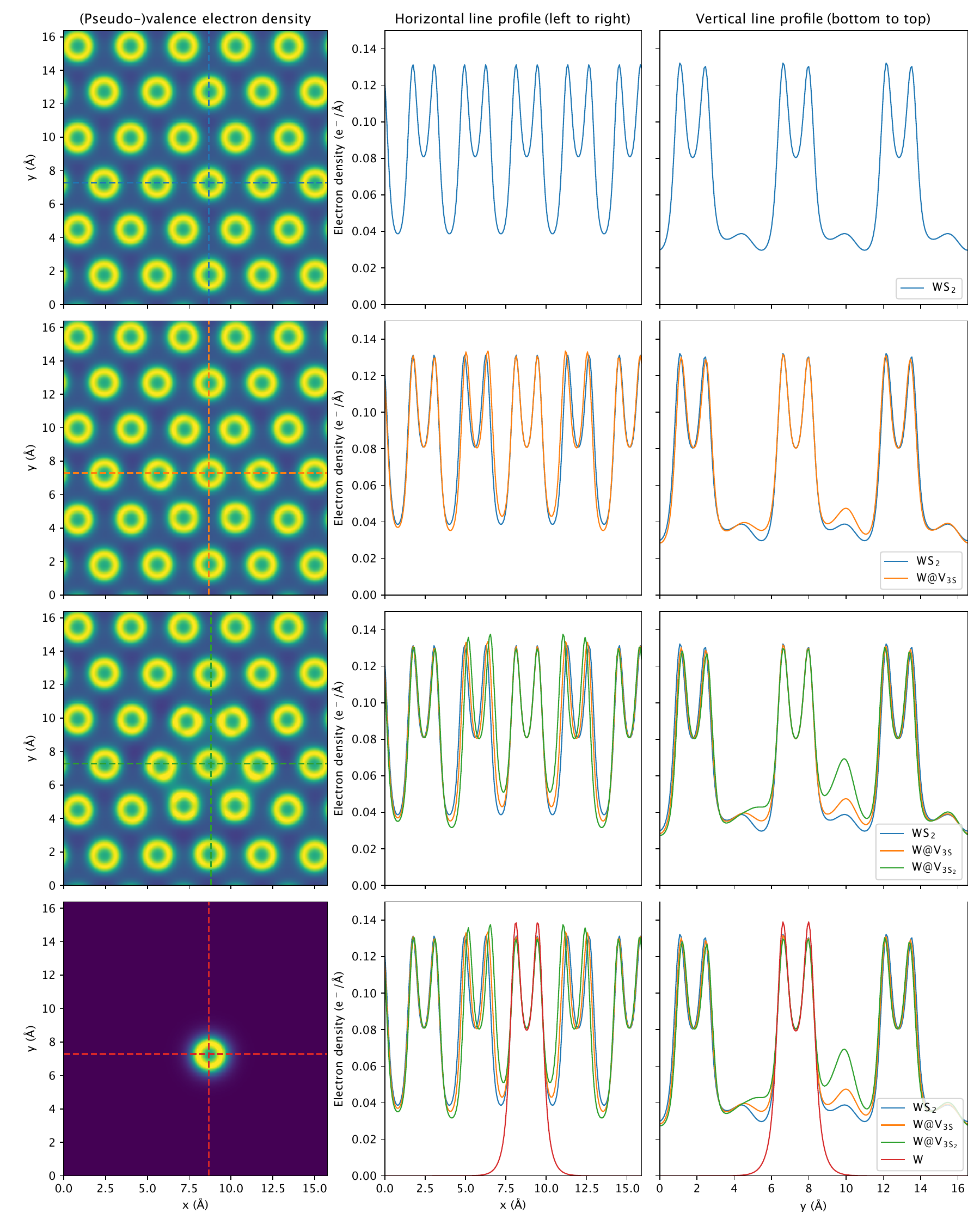}
 \includegraphics[width=0.55\textwidth]{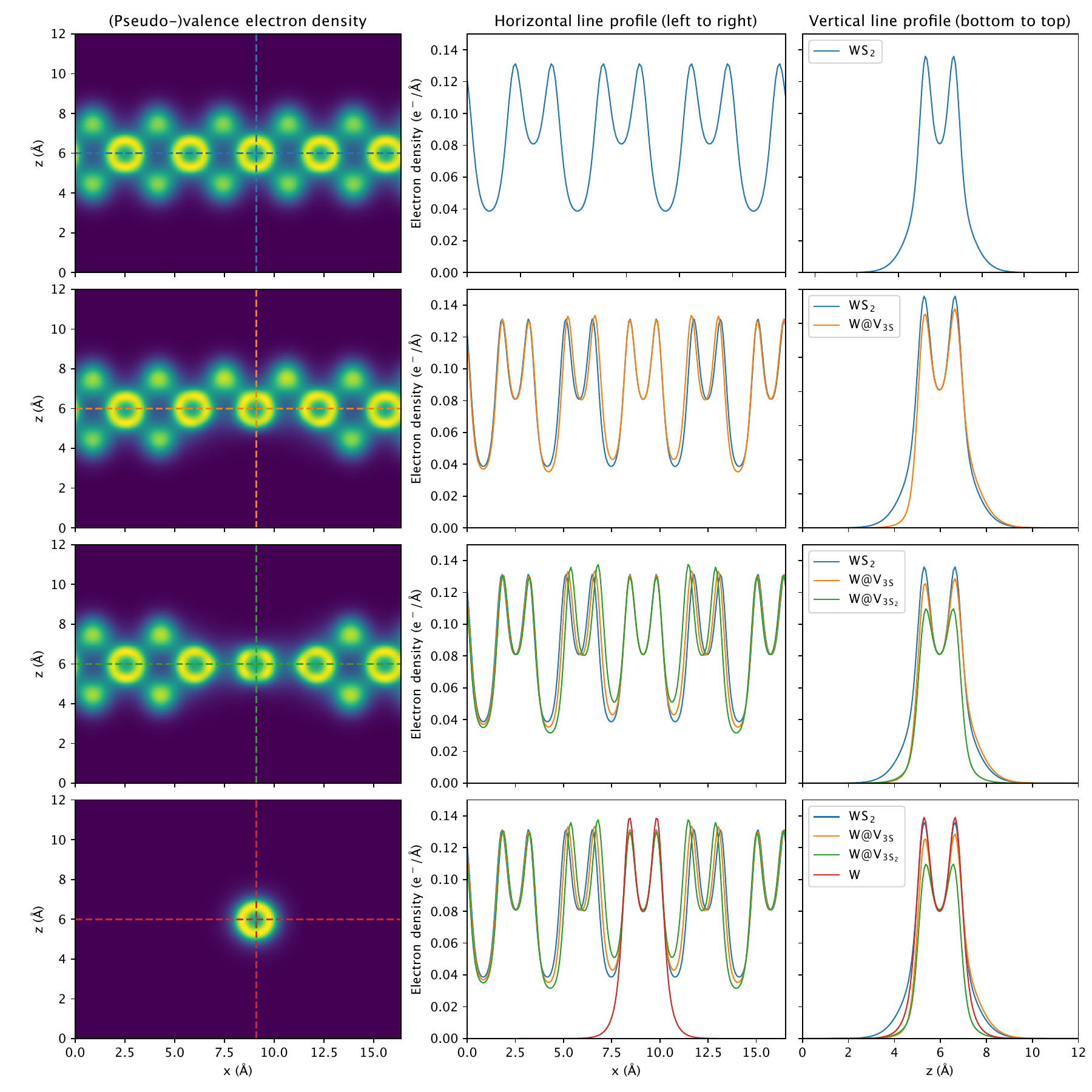}%
 \caption{\label{valence} \textbf{Pseudo valence electron density of (defective) WS\textsubscript{2} in top view (left) and side view (right).} From the 3D pseudo valence densities, the maps (first and third column) show a 2D slice (0.07~\AA~thickness) at the height of the W atoms (second and fourth column). The density of defects is increased from top to bottom. The valence electron density is decreasing at the W site close to the defect, explaining the phase increase.}
 \end{figure*}

SFig.~\ref{diff} shows the IAM and DFT projected potentials and their differences for the same defect configurations as in SFig.~\ref{diff1}. The same trend as in the SSB images is observed: as the vacancy density increases (left to right), the DFT potential increases at the W site next to the S vacancies compared to the IAM potential. This demonstrates that the phases are directly related to the potentials and the changes in them are caused by bonding and charge transfer.
 
To elucidate the origin of the increase of the phase at the W site that occurs with the increase in the density of vacancies around it, we calculated the 3D pseudo valence electron density for pristine WS\textsubscript{2}, V$_\mathrm{3S}$,  V$_{\mathrm{3S}_2}$, and an isolated W atom. From the 3D volume, we calculated top and side view 2D electron density maps of a single slice of 0.07~\AA\ thickness along the WS$_2$ sheet at the position of the W atom adjacent to the vacancies, as shown in SFig.~\ref{valence}. In this thin slice centered on the W layer of atoms, the electron density of the S atoms is not visibly apparent as viewed from the top. In both top and and side view electron density maps, the electron density can be seen to shift away from W site as vacancies are introduced with the V$_{\mathrm{3S}_2}$ configuration having the lowest density of valance electrons around the W@S\textsubscript{vac} and the greatest increase between it and the neighboring W atoms. Note that the completely isolated W atom has the greatest concentration of valance electron density around it. Therefore it is not simply the absence of S atoms around the a W atom that reduces the concentration of valence electrons around a W@S\textsubscript{vac}, but rather the charge transfer involved in a change in the bonding as the vacancies are introduced to the pristine structure. Compared to the isolated W atom, the electron density shifts away from the W atom when it is placed in the pristine structure. The electron density then shifts further away from the W atom as vacancies are introduced around it and the bonds essentially extend across longer distances. 
This is in agreement with the phase contrast we observe at different vacancy densities: the lower screening coming from the reduction of the electron density at a W site as the amount of vacancies increases also increases the  phase at the W@S\textsubscript{vac}. Thus the phase of the W@S\textsubscript{vac} is greatest for the V$_{\mathrm{3S}_2}$ configuration.
%


\end{document}